\documentclass[aps,pra,twocolumn,amsmath,amssymb,groupedaddress,longbibliography]{revtex4-2}
\usepackage{graphicx}
\usepackage{dcolumn}
\usepackage{bm}
\usepackage[pdfstartview=FitH, CJKbookmarks=true, bookmarksnumbered=true, bookmarksopen=true, colorlinks=true, pdfborder=001, citecolor=blue, linkcolor=blue, linktocpage=true] {hyperref}

\begin{document}

\preprint{APS/123-QED}
\title{Correlated topological pumping of interacting bosons assisted by Bloch oscillations}

\author{Wenjie Liu $^{1}$}

\author{Shi Hu $^{3}$}

\author{Li Zhang $^{1}$}

\author{Yongguan Ke $^{1}$}
\altaffiliation{Email: keyg@mail2.sysu.edu.cn}

\author{Chaohong Lee $^{1,2}$}
\altaffiliation{Email: lichaoh2@mail.sysu.edu.cn}

\affiliation{$^{1}$Guangdong Provincial Key Laboratory of Quantum Metrology and Sensing $\&$ School of Physics and Astronomy, Sun Yat-Sen University (Zhuhai Campus), Zhuhai 519082, China}

\affiliation{$^{2}$State Key Laboratory of Optoelectronic Materials and Technologies, Sun Yat-Sen University (Guangzhou Campus), Guangzhou 510275, China}

\affiliation{$^{3}$School of Photoelectric Engineering, Guangdong Polytechnic Normal University, Guangzhou 510665, China}

\date{\today}

\begin{abstract}

Thouless pumping, not only achieving quantized transport but also immune to moderate disorder, has attracted growing attention in both experiments and theories.
Here, we explore how particle-particle interactions affect topological transport in a periodically-modulated and tilted optical lattice.
Not limited to wannier states, our scheme ensures a dispersionless quantized transport even for initial Gaussian-like wave packets of interacting bosons which do not uniformly occupy a given band.
This is because the tilting potential leads to Bloch oscillations uniformly sampling the Berry curvatures over the entire Brillouin zone.
The interplay among on-site potential difference, tunneling rate and interactions contributes to the topological transport of bound and scattering states and the topologically resonant tunnelings.
Our study deepens the understanding of correlation effects on topological states, and provides a feasible way for detecting topological properties in interacting systems.

\end{abstract}

\maketitle

\section{Introduction}\label{Chap5Sec2}

Thouless pumping~\cite{1983PhysRevB276083,Niu1984}, a quantized transport in periodically modulated systems, can be viewed as a dynamical version of integer quantum Hall effect.
In a Thouless pump, when a uniform band occupation and adiabatic cyclic modulation are satisfied, the Chern number of the occupied band can be identified by the displacement per driven period.
Due to potential applications such as current standard, quantum state transfer and entanglement generation, Thouless pumping has been theorized~\cite{2013RevModPhys851421,2016LaserPhotonicsRev,2019PhysRevB100064302,2020PhysRevA101052323,2021PhysRevA104063315,2022SciPostPhys122022}, and observed in ultracold atom~\cite{2016NatPhy12350,2016NatPhy12296,2018nature55355,2021NatPhys17844}, photon~\cite{2012PhysRevLett109106402,2018Nature55359,2020Light9178} and spin~\cite{2018PhysRevLett120120501}systems.
However, if initial state is restricted to a single momentum state, a nonquantized geometric pumping has been observed in a Bose-Einstein Condensate~\cite{2016PhysRevLett116200402}.
Recently, a new topological pumping assisted by Bloch osillations is proposed, in which adding a tilting potential recovers a quantized transport of a single momentum state~\cite{2020PhysRevResearch2033143}.
Owing to the tilt potential, two main obstacles in Thouless pumping, initial-state preparation with a uniform band occupation and wave-packet dispersion, are solved.

Since particle-particle interaction is ubiquitous and inevitable, Thouless pumping has been naturally generalized to interacting systems~\cite{2016PhysRevLett117213603,2017PhysRevA95063630,2018PhysRevB98245148,2020PhysRevA101023620,2021nature59663}.
Relying on co-translational symmetry, Thouless pumping of a multiparticle Wannier state is related to the Chern number of a given multiparticle Bloch band.
For few particles, the interactions support the Thouless pumping of bound states in which particles are transport as a whole~\cite{2016PhysRevLett117213603,2020PhysRevA101023620}, and topologically resonant tunneling in which particles are transported one by one~\cite{2017PhysRevA95063630}.
For many particles, the quantized topological pumping may occur in the Mott-insulating regime with one boson per unit cell~\cite{2018PhysRevB98245148} and break down due to the vanishing many-body energy gap.
On the other hand, the onsite interactions can be treated as Kerr nonlinearity in the mean field approximation.
Moderate nonlinearity supports quantized transport of a solition while strong nonlinearity makes the soliton localized, which has been observed in curved waveguide arrays~\cite{2021nature59663,2022PhysRevLett128154101}.
Notice that a tilting potential has some advantages in topological pumping of a single particle.
It is of great interest how the new topological pumping assisted by Bloch oscillations is affected by particle-particle interaction.

\begin{figure}[htp]
\center
\includegraphics[width=0.45\textwidth]{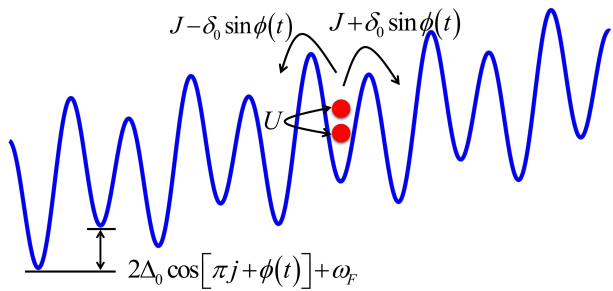}
\caption{Schematic of two bosons in a periodically driven superlattice under a tilted linear potential.
The nearest-neighbor tunneling and on-site potential are driven via the phase $\phi(t)$.
The on-site interaction is labelled by $U$ and the tilt is $\omega_F$.
\label{fig:schematic}}
\end{figure}

In this paper, we study topological transport of interacting bosons in a periodically driven tilted optical superlattice, as depicted in Fig.~\ref{fig:schematic}.
In a rotating-wave framework, the tilting potential is transferred to the role of linearly varying phase in the tunneling rate, and co-translational symmetry is recovered.
With the help of multi-particle Bloch theorem, the energy bands consist of scattering-state bands and bound-state bands in the strong interaction region.
For an initial bound state with an arbitrary center-of-mass momentum, we find that the center-of-mass displacement in a overall period is nearly quantized and independent of the initial quasi-momentum, in contrast to non-quantized displacement in the absence of tilting potential.
The quantization is related to a reduced Chern number defined as one-dimensional time integral of Berry curvature, which can be viewed a perfect approximation of two-body Chern number.
We derive an effective single-particle model for the bound states by the many-body degenerate perturbation theory, and find the reduced Chern number of the single-particle model is consistent with the one mentioned above.
The reduced Chern number can be used to precisely identify the boundary of topological phase transitions.
For the scattering states, when the two particles are far away from each other and occupy the topologically-nontrivial scattering-state bands, the two particles as a Fock state can be  transferred as a whole.
However, when the two particles occupy the topologically-trivial scattering-state bands,
the two particles can be independently shifted toward each other and come back to the initial state after crossing each other.
When on-site potential difference between two neighboring sites matches the interaction, we find topologically resonant tunnelings occur where two particles move one by one.
Surprisingly, the center-of-mass momentum is linearly and periodically varied as a function of time.

This paper is organized as follows.
In Sec.~\ref{Chap5Sec2}, we introduce an interacting Rice-Mele model in a tilted optical lattice.
In Sec.~\ref{Chap5Sec3}, under the guidance of effective single-particle model, we clarify the topological pumping of bound states assisted by Bloch oscillations.
In Sec.~\ref{TPSS}, we show the scattering states can be topologically transported as a whole or maintain the same after certain cycles, depending on the initial positions of the two particles.
In Sec.~\ref{Chap5Sec4}, we discuss the topologically resonant tunnelings assisted by Bloch oscillations when on-site potential difference between neighboring sites matches the interaction.
In Sec.~\ref{Chap5Sec6}, we give a brief summary and discussion.

\section{An interacting Rice-Mele model in a tilted optical lattice} \label{Chap5Sec2}

We consider an interacting Rice-Mele model with a tilted potential, described by the following Hamiltonian
\begin{equation}\label{Chap5RiceMeleModel}
\begin{aligned}
&\hat{H}(t)= \sum_{j}\left\{\left\{J+\delta_{0} \sin [\pi j+\phi(t)]\right\} \hat{a}_{j}^{\dagger} \hat{a}_{j+1}+\mathrm{H.c.}\right\} \\
&+\frac{U}{2} \sum_{j} \hat{n}_{j}\left(\hat{n}_{j}-1\right)+\sum_{j}\left\{\Delta_{0} \cos [\pi j+\phi(t)]+\omega_{F} j\right\} \hat{n}_{j}.
\end{aligned}
\end{equation}
Here, $\hat{a}_{j}^{\dagger}$($\hat{a}_{j}$) creates (annihilates) a boson at site $j$ with the atom number operator $\hat{n}_{j}=\hat{a}_{j}^{\dagger} \hat{a}_{j}$.
For simplicity, we set the Planck constant $\hbar=1$.
$J$ is the tunneling constant, $\delta_{0}$ and $\Delta_{0}$ respectively represent the modulation amplitudes of tunneling strength and on-site potential.
$U$ denotes the interaction which can be tuned by Feshbach resonance~\cite{2004PhysRevLett92160406,2010nature4641165}.
A double-well optical superlattice can be formed by a superposition of a short-wavelength optical lattice $V_{s}(t)=-V_{s} \cos ^{2}(\pi x/d )$ with a period $d$ and a long-wavelength optical lattice $V_{l}(t)=-V_{l} \cos ^{2}[\pi x /(2d) -\phi(t) / 2]$ with a period $2d$.
The relative phase $\phi(t)=\phi_{0}+\omega t$ has been experimentally realized~\cite{2016NatPhy12350,2016NatPhy12296}.

$\omega_{F}$ is a tilt formed by applying a gradient magnetic field or by placing the lattice along the gravitational field direction.
After applying additional tilts along one~\cite{2020PhysRevLett125236401,2020PhysRevA102063316} or two~\cite{2021PhysRevB104104314} directions in two-dimensional systems, the shifts of wave-packet centroid are also proposed to dynamically detect band topology.
The tilt $\omega_{F}$ forms an on-site potential difference between two neighboring sites, which breaks the lattice translational invariance.
To obtain the energy band under periodic boundary conditions, we apply a time-dependent unitary transformation $\hat{U}=\exp({i \sum_{j} j \omega_{F} t \hat{n}_{j}})$ to the Hamiltonian~\eqref{Chap5RiceMeleModel}, the lattice translational invariance is recovered in the rotating frame given by
\begin{equation}\label{Chap5RotatingFrame}
\begin{aligned}
&\hat{H}_{\text {rot }}(t)=\frac{U}{2} \sum_{j} \hat{n}_{j}\left(\hat{n}_{j}-1\right)+\sum_{j}\left\{\Delta_{0} \cos [\pi j+\phi(t)]\right\} \hat{n}_{j} \\
&+\sum_{j}\left\{\left\{J+\delta_{0} \sin [\pi j+\phi(t)]\right\} e^{-i \omega_{F} t} \hat{a}_{j}^{\dagger} \hat{a}_{j+1}+\mathrm{H.c.}\right\}.
\end{aligned}
\end{equation}
Obviously, the tilted potential in the Hamiltonian~\eqref{Chap5RiceMeleModel} amounts to a time-dependent phase factor in the tunneling term in the rotating frame.
The Hamiltonian~\eqref{Chap5RotatingFrame} involves in two frequencies where one is the modulation frequency $\omega$ depending on the parameter modulation period $T_{m}=2 \pi / \omega$ and the other is the tilt frequency $\omega_{F}$ depending on the tilt period $T_{F}=2\pi / \omega_{F}$.
%
$p$ and $q$ are chosen as coprime numbers, and a rational number is defined as $\omega_{F} / \omega=p / q$.
In the following sections, the evolution time $T_{\mathrm{tot}}$ is chosen as the common multiple of periods $T_{m}$ and $T_{F}$ with $T_{\mathrm{tot}}=nq T_{m}$, $n=1,2,3,...$.

We consider $N$ interacting bosons in an optical superlattice consisting of $L$ cells with $d$ sites per cell.
For Hamiltonian~\eqref{Chap5RotatingFrame}, particles as a whole satisfy a cotranslational invariance, that is, the system remains invariant as long as all particles as a whole shift integer cells~\cite{2003PhysRevE68056213,2017PhysRevA95063630}.
The quasimomentum of the center of mass, as a good quantum number, contributes to the multiparticle Bloch bands by solving the eigenequation
\begin{equation}\label{Chap5EigenFuction}
\hat{H}_{\text {rot }}(k)\left|\psi_{m}(k)\right\rangle=E_{m}(k)\left|\psi_{m}(k)\right\rangle.
\end{equation}
Here, $\left|\psi_{m}(k)\right\rangle$ represents the multiparticle Bloch state with quasimomentum $k$ in the $m$th multiparticle Bloch bands whose corresponding eigenvalue is $E_{m}(k)$.
Besides the cotranslational invariance in the optical superlattice, the Hamiltonian~\eqref{Chap5RotatingFrame} possesses a periodicity in time domain with $\hat{H}_{\mathrm{rot}}\left(t+T_{\mathrm{tot}}\right)=\hat{H}_{\mathrm{rot}}(t)$ with a overall period $T_{\mathrm{tot}}$.
Combining the periodic parameters $t$ and $k$, we can construct a closed surface where $t$ serves as the quasimomentum in the second dimension~\cite{2012PhysRevLett108220401,2013PhysRevLett110075303,2020PhysRevResearch2033143}.
Referring to real two-dimensional periodic systems~\cite{1982PhysRevLett49405}, we define the Chern number in two-dimensional closed surface as
\begin{equation}\label{Chap5ChernNumber}
C_{m}=\frac{1}{2 \pi} \int_{0}^{2 \pi / d} d k \int_{0}^{T_{\mathrm{tot}}} d t \mathcal{F}_{m}(k, t)
\end{equation}
with the Berry curvature of the $m$th band $\mathcal{F}_{m}=i\left(\left\langle\partial_{t} \psi_{m} \mid \partial_{k} \psi_{m}\right\rangle-\left\langle\partial_{k} \psi_{m} \mid \partial_{t} \psi_{m}\right\rangle\right)$.

Given $\left[\hat{H}_{\text {rot}}, \sum_{j} \hat{n}_{j}\right]=0$, subspaces with different particle numbers are decoupled and the particle number is conserved.
We mainly analyze the topological properties of two interacting bosons in the Rice-Mele model where the system is confined in two-boson basis $\left\{\left|l_{1} l_{2}\right\rangle=\left(1+\delta_{l_1l_2}\right)^{-1 / 2} \hat{a}_{l_{1}}^{\dagger} \hat{a}_{l_{2}}^{\dagger}|\mathbf{0}\rangle\right\}$ with $1 \leq l_{1} \leq l_{2} \leq L_{t}$ and
$L_{t}=dL$ being the system size.
After introducing $C_{l_{1} l_{2}}=\left\langle\mathbf{0}\left|\hat{a}_{l_{2}} \hat{a}_{l_{1}}\right| \boldsymbol{\psi}\right\rangle$, an arbitrary two-boson state can be expanded as
$|\boldsymbol{\psi}\rangle=\sum_{l_{1} \leq l_{2}} \psi_{l_{1} l_{2}}\left|l_{1} l_{2}\right\rangle$
where $\psi_{l_{1} l_{2}}=C_{l_{1} l_{2}}(1+\delta_{l_1l_2})^{-1/2}$.
Further, the eigenequation $\hat{H}_{\mathrm{rot}}|\boldsymbol{\psi}\rangle=E|\boldsymbol{\psi}\rangle$ turns to
\begin{widetext}
\begin{equation}
\begin{aligned}
&E C_{l_{1} l_{2}}=\left\{J+\delta_{0} \sin \left[\pi l_{1}+\phi(t)\right]\right\} e^{-i \omega_{F} t} C_{l_{1}+1, l_{2}}+\left\{J+\delta_{0} \sin \left[\pi l_{2}+\phi(t)\right]\right\} e^{-i \omega_{F} t} C_{l_{1}, l_{2}+1} \\
&+\left\{J+\delta_{0} \sin \left[\pi\left(l_{1}-1\right)+\phi(t)\right]\right\} e^{i \omega_{F} t} C_{l_{1}-1, l_{2}}+\left\{J+\delta_{0} \sin \left[\pi\left(l_{2}-1\right)+\phi(t)\right]\right\} e^{i \omega_{F} t} C_{l_{1}, l_{2}-1} \\
&+\left\{\Delta_{0} \cos \left[\pi l_{1}+\phi(t)\right]+\Delta_{0} \cos \left[\pi l_{2}+\phi(t)\right]\right\} C_{l_{1} l_{2}}+U C_{l_{1} l_{2}} \delta_{l_{1} l_{2}} .
\end{aligned}
\end{equation}
\end{widetext}

\begin{figure}[htp]
\center
\includegraphics[width=0.45\textwidth]{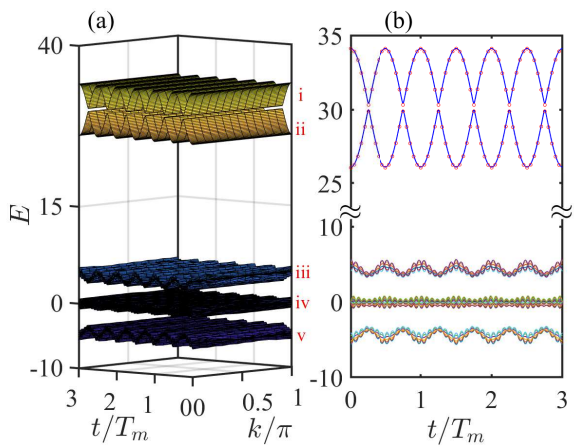}
\caption{Two-boson energy spectrum.
(a) The 3D view of Bloch bands in the parameter space ($k$, $t$).
(b) The time evolution of energies $E(t)$ for Bloch states with $k=0$.
The red circles are the eigenenergies of the effective single-particle model~\eqref{Chap5EffectiveModel}.
The parameters are chosen as $J=-1$, $\delta_{0}=0.8$, $\Delta_{0}=2$, $U=30$, $\omega=0.005$, $\phi_{0}=0$, $\omega_{F} / \omega=10 / 3$, $L_t=26$.
\label{fig:boundspectrum}}
\end{figure}

\section{Topological pumping of bound states assisted by Bloch oscillations} \label{Chap5Sec3}

In this section we discuss the quantized topological pumping of two-boson bound states assisted by Bloch oscillations, where bosons tend to stay at the same site.
By solving the two-particle eigenequation~\eqref{Chap5EigenFuction}, we obtain two-particle Bloch bands; see Fig.~\ref{fig:boundspectrum}(a).
The parameters are chosen as $J=-1$, $\delta_{0}=0.8$, $\Delta_{0}=2$, $U=30$, $\omega=0.005$, $\phi_{0}=0$, $\omega_{F} / \omega=10 / 3$, $L_t=26$.
The two-particle energy spectrum consists of five isolated bands ordered for decreasing values of the energy marked with bands (i-v), of which the bands (i-ii) correspond to bound states and the rest bands belong to scattering states.
According to the definition~\eqref{Chap5ChernNumber}, the Chern numbers of multiparticle Bloch bands are calculated: $C_{\rm{i}}=-3$, $C_{\rm{ii}}=3$, $C_{\rm{iii}}=-3$, $C_{\rm{iv}}=0$ and $C_{\rm{v}}=3$.
The corresponding bulk-boundary correspondence under open boundary condition is presented in Appendix~\ref{BBC}.
Fig.~\ref{fig:boundspectrum}(b) shows the change of energies for Bloch states with $k=0$, where the band gaps remain open as time evolves.

\begin{figure}[htp]
\center
\includegraphics[width=0.35\textwidth]{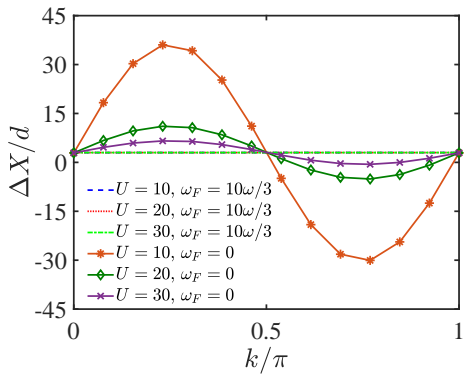}
\caption{The center-of-mass displacements of multiparticle Bloch states with different quasimomenta $k$ in terms of the semiclassical formula~\eqref{Chap5Semiclassical} for different $U$ and $\omega_F$.
The other parameters are chosen as $J=-1$, $\delta_{0}=0.8$, $\Delta_{0}=2$, $\omega=0.005$, $\phi_{0}=0$ and $L_t=26$.}
\label{fig:boundsemifunction}
\end{figure}

The adiabatic transport theorem indicates the velocity for a state with momentum $k$ in the $m$th band is written to first order as~\cite{2010RevModPhys821959}
\begin{equation}\label{Chap5GroupV}
v_m(k, t)=\frac{\partial \varepsilon_{m}(k, t)}{\hbar \partial k}+\mathcal{F}_{m}(k, t).
\end{equation}
After inputting a Bloch state, the pumping distance at the moment $\tau$ is determined by a semiclassical formula
\begin{equation}\label{Chap5Semiclassical}
\Delta X(\tau)=\int_{0}^{\tau} v_m(k, t) d t.
\end{equation}
Without loss of generality, we pay attention on the band (ii) of the two-boson Bloch bands.
Fig.~\ref{fig:boundsemifunction} displays the center-of-mass displacements of multiparticle Bloch states with different quasimomenta in the band (ii) after an evolution time $T_{\mathrm{tot}}=3T_m$.
When $\omega_F=0$, similar to a geometric pumping~\cite{2016PhysRevLett116200402}, the center-of-mass displacements of Bloch states are associated with the quasimomentum $k$, whose amplitudes are significantly reduced by interaction.
In the presence of $\omega_F$, $\Delta X\left(3T_ {m}\right)/ d $ for each $k$ are quite close to the band Chern number ($C_{\rm{ii}}=3$), independent of the values of interaction.
It implies a quantized transport can be realized even for initial Bloch states.

The multiparticle Bloch state is written in the two-particle basis as $|\boldsymbol{\psi}\rangle=\sum_{l_{1} \leq l_{2}} \psi_{l_{1} l_{2}}\left(k_{0},t_{0}\right)\left|l_{1} l_{2}\right\rangle$ with the probability amplitude $\psi_{l_{1} l_{2}}\left(k_{0},t_{0}\right)$.
Since multiparticle Bloch states independently from $k$ can realize quantized transports, it is reasonable to predict that a quantized transport can be reached with an arbitrary initial wave packet prepared on a focused band.
In order to verify this, two typical initial states are considered.
One is a strongly localized initial state in real space, e.g. $\left|\boldsymbol{\psi}\left(t_{0}\right)\right\rangle=\left|L_t/2, L_t/2\right\rangle$.
The multiparticle Bloch state with quasimomentum $k_ {0}=0$ in the band (ii) is constructed into Gaussian wave packet as another type of initial state, e.g.
\begin{equation}\label{Chap5GaussianWave}
|\boldsymbol{\psi}(t_{0})\rangle=\sum_{l_{1} \leq l_{2}} e^{-\frac{\left(l_{1}-l_{0}\right)^{2}+\left(l_{2}-l_{0}\right)^{2}}{4 \sigma^2}} \psi_{l_{1}l_{2}}\left(k_{0},t_{0}\right)\left|l_{1} l_{2}\right\rangle.
\end{equation}
Here $\sigma$ and $l_{0}$ are the width and center-of-mass position of the initial wave packet, respectively.
Such Gaussian wave packet~\eqref{Chap5GaussianWave} can be prepared by adding an additional harmonic potential.
The evolved state $|\boldsymbol{\psi}(t)\rangle=\sum_{l_{1} \leq l_{2}} \psi_{l_{1} l_{2}}(t)\left|l_{1} l_{2}\right\rangle$ is determined by $|\boldsymbol{\psi}(t)\rangle=\mathcal{T} \exp \left\{-i \int_{t_{0}}^{t} \hat{H}(t)d t\right\}\left|\boldsymbol{\psi}\left(t_{0}\right)\right\rangle$
with the time-ordering operator $\mathcal{T}$.
We extract the center-of-mass displacement
\begin{equation}
\Delta X(t)=X(t)-X(0)
\end{equation}
from the particle density distribution
\begin{equation}
n_{j}(t)=\left\langle\boldsymbol{\psi}(t)\left|\hat{n}_{j}\right| \boldsymbol{\psi}(t)\right\rangle
\end{equation}
with $X(t)=\sum_{l_{1} \leq l_{2}} \frac{l_{1}+l_{2}}{2}\left|\psi_{l_{1} l_{2}}(t)\right|^{2}$.

\begin{figure*}[htp]
\center
\includegraphics[width=0.9\textwidth]{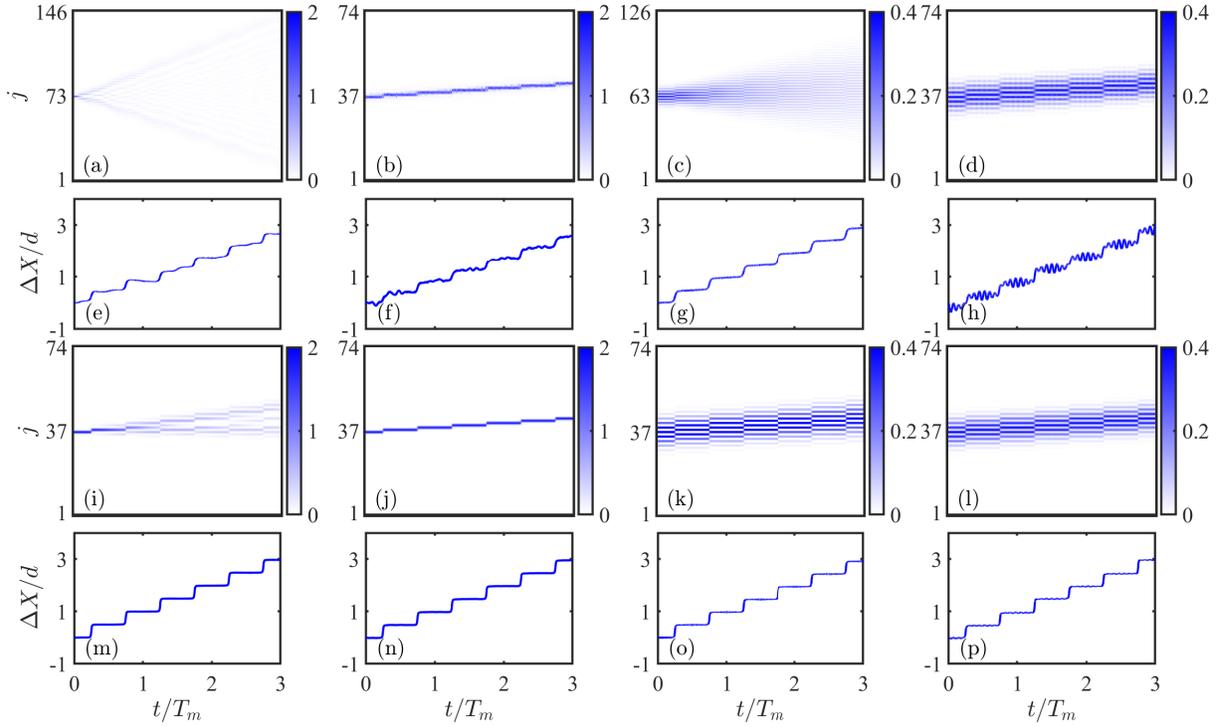}
\caption{The time evolution of particle density distribution and its corresponding center-of-mass displacement of the bound-state dynamics.
The role of tilt is analyzed: $\omega_F=0$ (the first and third columns) and $\omega_F=10\omega/3$ (the second and fourth columns).
Two types of initial states are discussed: localized Fock states (the first and second columns) and extended Gaussian states (the third and fourth columns).
Different interaction strengths are examined: $U=10$ (the first and second rows) and $U=30$ (the third and fourth rows).
The other parameters are chosen as $J=-1$, $\delta_{0}=0.8$, $\Delta_{0}=2$, $\omega=0.005$, $\phi_{0}=0$ and $T_{\mathrm{tot}}=q T_{m}=3 T_{m}$.}
\label{fig:boundtransport}
\end{figure*}

By numerically calculating the wave-packet dynamics of different initial states and interaction strengths, we evaluate the effect of tilt in Fig.~\ref{fig:boundtransport}.
Relying on the probability of initial states projecting onto band (ii), the motion of real-space wave packet undergoes distinct dynamics.
In the weak interaction regime, the tightly localized initial states actually spread in bands (i) and (ii) where fidelity between localized initial states and band (ii) increases as interaction increases.
For $U=10$, the fidelity between initial Fock state and band (ii) is $0.858$, accompanied by the nonquantized centroid displacements $2.644d$ ($\omega_F=0$) and $2.607d$ ($\omega_F=10\omega/3$).
Despite the lack of tilt, the centroid displacements of initial Gaussian wave packet gradually reach to band topology $C_{\rm{ii}}=3$ as interaction increases.
In agreement with semiclassical results in Fig.~\ref{fig:boundsemifunction}, this behavior is directly attributed to the reduced group velocity by the interaction.
We find that, the role of a tilt ensures the dispersionless quantized centroid displacements once the initial states are prepared at a specific band following adiabatic evolutions,
while for $\omega_F=0$ the dispersion quantized centroid displacements are obtained with the initial states uniformly filling at a
specific band.
Our results confirm that the initial states that lack a uniform band occupation can support quantized transport when applied to a tilt.
At the same time, the wave packet oscillates in real space with a slight width due to the reduced tunneling strength by strong interaction, whose effect of tilt is obviously exhibited in quasimomentum space in Appendix~\ref{DDIQS}.

\subsection{Effective single-particle model} \label{Chap5Sec31}

To have a better understanding of the bound-state transport mechanism, we take the interaction term as the domination and the other terms as the perturbation, and employ the many-body degenerate perturbation theory to analytically derive an effective single-particle model of bound states.
The effective single-particle model can greatly describe the topological properties of bound-state bands,
where eigenvalues, eigenstates and topological invariants of the bound-state bands can be solved analytically.
Consequently, the reduced Chern number of the multiparticle Bloch bands can be defined.

The Hamiltonian~\eqref{Chap5RiceMeleModel} is divided into two parts, $\hat{H}(t)=\hat{H}_{0}+\hat{H}_{\text {int }}$ with
\begin{equation}
\hat{H}_{\text {int }}=\frac{U}{2} \sum_{j} \hat{n}_{j}\left(\hat{n}_{j}-1\right)
\end{equation}
as the interaction term and
\begin{equation}
\begin{aligned}
\hat{H}_{0}=& \sum_{j}\left\{\left\{J+\delta_{0} \sin [\pi j+\phi(t)]\right\} \hat{a}_{j}^{\dagger} \hat{a}_{j+1}+\rm{H . c .}\right\} \\
&+\sum_{j}\left\{\Delta_{0} \cos [\pi j+\phi(t)]+ \omega_{F} j\right\} \hat{n}_{j}
\end{aligned}
\end{equation}
as the noninteraction term.
In the strongly interacting regime $U \gg\left(J, \delta_{0}, \Delta_{0}, \omega_{F}\right)$, $\hat{H}_{0}$ is treated as a perturbation of $\hat{H}_{\mathrm{int}}$.
$\hat{H}_{\text {int }}$ includes two degenerate subspaces $\mathcal{U}$ and $\mathcal{V}$.
The subspace $\mathcal{U}$ consists of bound states with two bosons at the same site, that is, $\mathcal{U}\equiv\left\{|2\rangle_{j}\right\}$, and the degenerate energy is $E_{j}=U$.
The subspace $\mathcal{V}$ consists of states with two bosons at different sites, that is, $\mathcal{V}\equiv\left\{|1\rangle_{j}|1\rangle_{k}\right\} $, and the degenerate energy is $E_{j k}=0$ with $j\neq k$.
We respectively define the projection operators into the subspaces $\mathcal{U}$ and $\mathcal{V}$ as
\begin{equation}
\hat{P}=\sum_{j}|2\rangle_{j}\langle 2|_{j}
\end{equation}
and
\begin{equation}
\hat{S}=\sum_{j \neq k} \frac{1}{E_{j}-E_{jk}}|1\rangle_{j}|1\rangle_{k}\langle 1|_{k}\langle 1|_{j}.
\end{equation}
By employing a second-order degenerate perturbation theory~\cite{Takahashi1977,BRAVYI20112793}, the effective Hamiltonian in the subspace $\mathcal{U}$ is given by
\begin{equation}
\hat{H}_{\mathrm{eff}}=\hat{h}_{0}+\hat{h}_{1}+\hat{h}_{2}=E_{j} \hat{P}+\hat{P} \hat{H}_{0} \hat{P}+\hat{P} \hat{H}_{0} \hat{S} \hat{H}_{0} \hat{P}.
\end{equation}
The zero-order term satisfies
\begin{equation}\label{Chap5ZeroTerm}
\hat{h}_{0}=E_{j} \hat{P}=U\sum_{j}|2\rangle_{j}\langle 2|_{j}.
\end{equation}
The first-order term is
\begin{equation}\label{Chap5FirstTerm}
\hat{h}_{1}=\hat{P} \hat{H}_{0} \hat{P}=\sum_{j}\left\{2\Delta_{0} \cos [\pi j+\phi(t)]+2 \omega_{F}j\right\}|2\rangle_{j}\langle 2|_{j}.
\end{equation}
And the second-order term becomes
\begin{equation}\label{Chap5SecondTerm}
\begin{aligned}
\hat{h}_{2} &=\hat{P} \hat{H}_{0} \hat{S} \hat{H}_{0} \hat{P}\\
=&\sum_{j}\frac{2\left\{J+\delta_{0} \sin [\pi j+\phi(t)] \right\}^2}{U}|2\rangle_{j}\langle 2|_{j+1}+\rm{H.c.}\\
&+\left[\frac{2\left(J-\delta_{0}\right)^{2}}{U}+\frac{2\left(J+\delta_{0}\right)^{2}}{U}\right] \sum_{j}|2\rangle_{j}\langle 2|_{j}
\end{aligned}
\end{equation}
Combining Eqs.~\eqref{Chap5ZeroTerm}, ~\eqref{Chap5FirstTerm} and~\eqref{Chap5SecondTerm}, the effective single-particle model yields
\begin{equation}\label{Chap5EffectiveModel}
\begin{aligned}
\hat{H}_{\mathrm{eff}}=& \sum_{j} \frac{2\left\{J+\delta_{0} \sin [\pi j+\phi(t)]\right\}^{2}}{U} \hat{b}_{j}^{\dagger} \hat{b}_{j+1}+\mathrm{H.c.} \\
&+\sum_{j}\left\{2 \Delta_{0} \cos [\pi j+\phi(t)]+2 \omega_{F} j\right\} \hat{b}_{j}^{\dagger} \hat{b}_{j}+\mathcal{C}.
\end{aligned}
\end{equation}
Here, $\hat{b}_{j}^{\dagger}$ represents the creation of two bosons at the site $j$ simultaneously, that is, $\hat{b}_{j}^{\dagger}|\mathbf{0}\rangle=|2\rangle_{j}$.
Given the conservation of the particle number, $\mathcal{C}$ is an energy constant with $\mathcal{C}=U+2\left(J-\delta_{0}\right)^{2}/U+2\left(J+\delta_{0}\right)^{2}/U$.
Similarly, the tilted potential breaks the lattice translational invariance which can be recovered in the rotating frame as
\begin{equation}\label{Chap5RotatorEffectiveModel}
\begin{aligned}
\hat{H}^{\rm{rot}}_{\text {eff }}=& \sum_{j} \frac{2\left\{J+\delta_{0} \sin [\pi j+\phi(t)]\right\}^{2}}{U} e^{-i 2 \omega_{F} t} \hat{b}_{j}^{\dagger} \hat{b}_{j+1} \\
&+\text { H.c. }+\sum_{j} 2 \Delta_{0} \cos [\pi j+\phi(t)] \hat{b}_{j}^{\dagger} \hat{b}_{j}+\mathcal{C}.
\end{aligned}
\end{equation}

\subsection{The reduced Chern number of multiparticle bound-state bands} \label{Chap5Sec32}

In order to obtain the eigenvalues and eigenstates of the bound-state bands in quasimomentum space, we take the Fourier transform for the effective single-particle model~\eqref{Chap5RotatorEffectiveModel} as
\begin{equation}\label{Chap5FFT}
\begin{array}{l}
\hat{b}_{2 j}^{\dagger}=\frac{1}{\sqrt{L}} \sum_{k} e^{-i k 2 j} \hat{b}_{k, \mathrm{e}}^{\dagger}, \\
\hat{b}_{2 j-1}^{\dagger}=\frac{1}{\sqrt{L}} \sum_{k} e^{-i k(2 j-1)} \hat{b}_{k, \mathrm{o}}^{\dagger}.
\end{array}
\end{equation}
Here $k$ is the quasimomentum and $L$ is the cell number.
$\mathrm{o}$($\mathrm{e}$) represents the odd (even) site.
After the Fourier transform~\eqref{Chap5FFT}, the Hamiltonian~\eqref{Chap5RotatorEffectiveModel} can be decomposed as $\hat{H}^{\rm{rot}}_{\text {eff }}(t)=\sum_{k} \hat{H}^{\rm{rot}}_{\text {eff }}(k,t)$.
Each $\hat{H}^{\rm{rot}}_{\text {eff }}(k,t)$ belongs to a two-level quantum system
\begin{equation}\label{Chap5TwoBandHamiltonian}
\hat{H}_{\text {eff }}^{\text {rot }}(k, t)=h_{x} \hat{\sigma}_{x}+h_{y} \hat{\sigma}_{y}+h_{z} \hat{\sigma}_{z}+\mathcal{C},
\end{equation}
where
\begin{equation}
\left(\begin{array}{l}
h_{x} \\
h_{y} \\
h_{z}
\end{array}\right)=\left(\begin{array}{c}
4 \frac{\left\{J^{2}+\delta_{0}^{2} \sin ^{2}[\phi(t)]\right\}}{U} \cos \left(k-2 \omega_{F} t\right) \\
-4 \frac{\left\{2 J \delta_{0} \sin [\phi(t)]\right\}}{U} \sin \left(k-2 \omega_{F} t\right) \\
2 \Delta_{0} \cos [\phi(t)]
\end{array}\right).
\end{equation}
By solving the eigenequation $\hat{H}_{\mathrm{eff}}^{\mathrm{rot}}(k, t)|u(k, t)\rangle=\varepsilon(k, t)|u(k, t)\rangle$, the eigenvalue is
\begin{equation}\label{Chap5TwoBoundBand}
\varepsilon_{\pm} =\pm \sqrt{h_{x}^{2}+h_{y}^{2}+h_{z}^{2}}+\mathcal{C}.
\end{equation}
The eigenstate satisfies
\begin{equation}
\left|u_{\pm}(k, t)\right\rangle=\left(\begin{array}{c}
\frac{h_{x}-i h_{y}}{\varepsilon_{\pm}-h_{z}} \\
1
\end{array}\right).
\end{equation}
Under the same parameters, the eigenvalue~\eqref{Chap5TwoBoundBand} is added in Fig.~\ref{fig:boundspectrum}(b) with red circles, which well agrees with the two bound-state energies.

Based on equivalent definition of the Berry curvature
\begin{equation}
\mathcal{F}_{m}(k, t)=-2 \operatorname{Im}\left[\sum_{m^{\prime} \neq m} \frac{\left\langle u_{m}\left|\partial_{k} \hat{H}\right| u_{m^{\prime}}\right\rangle\left\langle u_{m^{\prime}}\left|\partial_{t} \hat{H}\right| u_{m}\right\rangle}{\left(\varepsilon_{m}-\varepsilon_{m^{\prime}}\right)^{2}}\right]
\end{equation}
with $m=\pm$, we derive the Berry curvatures of two bands in the effective single-particle model~\eqref{Chap5RotatorEffectiveModel} as
\begin{widetext}
\begin{equation}\label{Chap5BerryCurvature1}
\mathcal{F}_{\pm}(k, t)=32 \frac{J \delta_{0} \omega \Delta_{0}}{U} \frac{\left\{J^{2}+\delta_{0}^{2} \sin ^{2}[\phi(t)]\right\}}{U} \frac{1-\cos ^{2}[\phi(t)] \cos ^{2}\left(k-2 \omega_{F} t\right)}{\left[\varepsilon_{\pm}(k, t)\right]^{3}}.
\end{equation}
\end{widetext}
The instantaneous eigenvalue $\varepsilon_{\pm}(k,t)$ as a periodic function of $k$ and $t$ ensures the integration of the dispersion velocity in the semiclassical formula~\eqref{Chap5Semiclassical} is equal to zero.
After an evolution time $T_{\rm{tot}}$, only the integration over the time of Berry curvature contributes to the center-of-mass displacement of the wave packet,
which is defined as the reduced Chern number
\begin{equation}\label{Chap5ReducedChernNumber}
C_{m, \mathrm{red}}\left(q T_{m}\right) \equiv \frac{\Delta X\left(q T_{m}\right)}{d}=\frac{1}{d} \int_{0}^{q T_{m}} \mathcal{F}_{m}(k, t) d t.
\end{equation}
The reduced Chern numbers obtained from Eq.~~\eqref{Chap5BerryCurvature1} and Eq.~\eqref{Chap5ReducedChernNumber} yield the values: $2.9604$($U=10$) and $2.9921$($U=30$).
The other parameters are chosen as $J=-1$, $\delta_{0}=0.8$, $\Delta_{0}=2$, $\omega=0.005$, $\omega_F=10\omega/3$, $\phi_{0}=0$ and $T_{\mathrm{tot}}=q T_{m}=3 T_{m}$.
Compared with band topology $C_{\rm{ii}}=3$, the slight deviation arises from the perturbation condition for effective model~\eqref{Chap5EffectiveModel}.
Given the validity of the effective model, the results become more accurate for stronger interaction.
It indicates the bound-state topological properties can be well described by the effective model for enough interaction strength.
In Appendix~\ref{appendixRC}, we have analytically derived a vital relationship $C_{m, \mathrm{red}}\left(q T_{m}\right)=C_m$ between the reduced Chern number and Chern number of the bound-state bands for $\omega_{F} / \omega \rightarrow \infty$.

\subsection{Detecting topological phase transitions by the reduced Chern number} \label{Chap5Sec5}

\begin{figure}[htp]
\center
\includegraphics[width=0.4\textwidth]{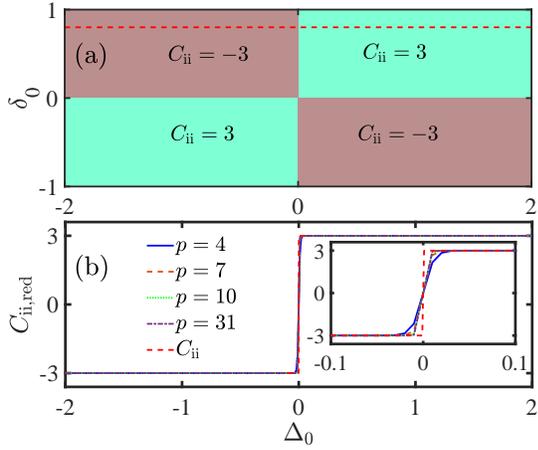}
\caption{(a) The Chern number $C_{\rm{ii}}$ of band (ii) with $\omega_{F}=0$.
(b) The detection of topological phase transition in terms of the Chern number $C_{\rm{ii}}$ with $\omega_F=0$ and reduced Chern number $C_{\rm{ii,red}}$ with $\omega_F=p\omega/q$.
The inset is an enlarged part in (b).
The other parameters are chosen as $J=-1$, $\omega=0.005$, $\phi_{0}=0$ and $T_{\mathrm{tot}}=q T_{m}=3 T_{m}$.
The system size is $L_t=58$.}
\label{fig:topologicalphase}
\end{figure}

The measurement of topological invariants plays an important role in understanding topological phases.
The band topology induces the quantized transport which may provide a useful way for detecting topological phase transitions in an interacting system.
By solving degenerate points of two bound-state bands, the topological phase boundary without a tilt is analytically obtained under the perturbation condition.
The system exists quantum criticality at $\Delta_0=0$ ($\delta_0=0$) regardless of $\delta_0$ ($\Delta_0$).
Except for the critical boundaries, the Chern number $C_{\rm{ii}}$ of the band (ii) is calculated in terms of Eq.~\eqref{Chap5ChernNumber}, and the corresponding topological phase diagram is shown in Fig.~\ref{fig:topologicalphase}(a).
The parameters are chosen as $J=-1$, $\omega=0.005$, $\omega_{F}=0$ and $\phi_{0}=0$.
The system size is $L_t=58$.
We focus on the topological phase transition from $C_{\rm{ii}}=-3$ to $C_{\rm{ii}}=3$ when $\Delta_0 $ takes value from $-2$ to $2$ at $\delta_0=0.8$, as shown in the red dashed lines in Figs.~\ref{fig:topologicalphase}(a) and (b).
The transition of Chern number is much sharp across the critical point $\Delta_0=0$.
when the tilt is present, the reduced Chern number $C_{\rm{ii}, \mathrm{red}}$ in the band (ii) is calculated by Eq.~\eqref{Chap5ReducedChernNumber}.
It is found that the reduced Chern number can detect the critical point more accurately as the tilt $\omega_F$ increases, as shown in Fig.~\ref{fig:topologicalphase}(b).
Therefore, the reduced Chern number offers a flexible way for the detection of topological phase transitions in interacting systems.

\section{Topological pumping of scattering states assisted by Bloch oscillations} \label{TPSS}

\begin{figure}[htp]
\center
\includegraphics[width=0.45\textwidth]{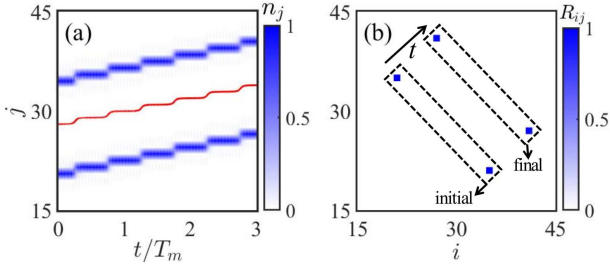}
\caption{Topological pumping of scattering states $\left|\boldsymbol{\psi}\left(t_{0}\right)\right\rangle=\left|21, 35\right\rangle$.
(a) The time evolution of the particle density distribution.
The center-of-mass position at each moment is showed with red line.
(b) The two-boson correlation at moments $t/T_m=0$ and $3$.
The parameters are chosen as $J=-1$, $\delta_{0}=0.8$, $\Delta_{0}=7$, $U=30$, $\omega=0.05$, $\phi_{0}=0$, $\omega_{F} / \omega=31 / 3$, $L_t=58$.
The evolution time is $T_{\mathrm{tot}}=3 T_{m}$.}
\label{fig:ss2135}
\end{figure}

The continuum band of scattering states is divided into three isolated cluster bands (iii)-(v).
At the initial moment ($t/T_m=0$), in band (v) two independent bosons mainly occupy the different odd sublattices;
in band (iv) one (the other) boson mainly occupies the odd (even) sublattice;
and in band (iii) two independent bosons mainly occupy the different even sublattices.
Under the appropriate parameters ($J=-1$, $\delta_{0}=0.8$, $\Delta_{0}=7$, $U=30$, $\omega=0.05$, $\phi_{0}=0$, $\omega_{F} / \omega=31 / 3$, $L_t=58$), initial state $\left|\boldsymbol{\psi}\left(t_{0}\right)\right\rangle=\left|21, 35\right\rangle$ ($\left|\boldsymbol{\psi}\left(t_{0}\right)\right\rangle=\left|23, 36\right\rangle$)
supports fidelity $0.9805$ ($0.9806$) with band (v) (band (iv)).
Fig.~\ref{fig:ss2135} show the dynamic evolution from the initial state $\left|\boldsymbol{\psi}\left(t_{0}\right)\right\rangle=\left|21, 35\right\rangle$ with an evolved time $T_{\mathrm{tot}}=3 T_{m}$.
Individual transport is revealed via the particle density distribution $n_{j}=\left\langle\boldsymbol{\psi}(t)\left|\hat{n}_{j}\right| \boldsymbol{\psi}(t)\right\rangle$ in Fig.~\ref{fig:ss2135}(a) and two-boson correlation $R_{i j}=\langle\boldsymbol{\psi}(t)|\hat{a}_{i}^{\dagger} \hat{a}_{j}^{\dagger} \hat{a}_{j} \hat{a}_{i}| \boldsymbol{\psi}(t)\rangle$ in Fig.~\ref{fig:ss2135}(b).
The time evolution of centroid position is added in Fig.~\ref{fig:ss2135}(a) with red line, which reflects the centroid shifts 2.9414 cells during the evolution process.
Similar to the behavior in Fig.~\ref{fig:ss2135}, initial two bosons mainly occupying in band (iii) will freely propagate in another direction due to $C=-3$.

\begin{figure}[htp]
\center
\includegraphics[width=0.45\textwidth]{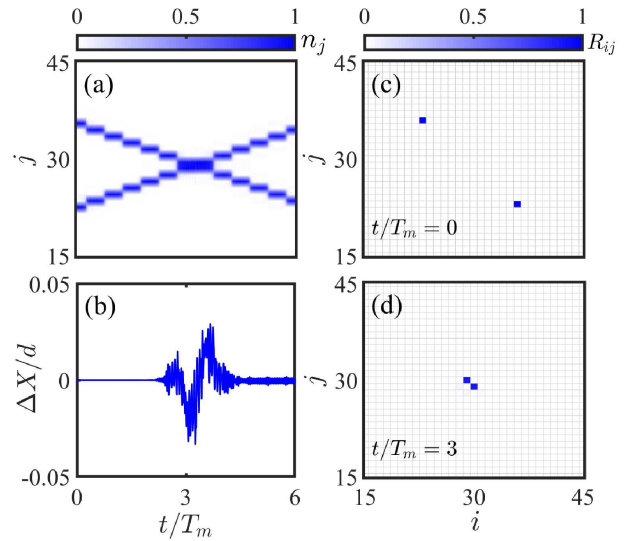}
\caption{Topological pumping of scattering states $\left|\boldsymbol{\psi}\left(t_{0}\right)\right\rangle=\left|23, 36\right\rangle$.
The time evolution of the particle density distribution (a) and center-of-mass displacement (b).
(c) and (d) are the two-boson correlations at moments $t/T_m=0$ and $t/T_m=3$.
The parameters are chosen as $J=-1$, $\delta_{0}=0.8$, $\Delta_{0}=7$, $U=30$, $\omega=0.05$, $\phi_{0}=0$, $\omega_{F} / \omega=31 / 3$, $L_t=58$.
The evolution time is $T_{\mathrm{tot}}=6 T_{m}$.}
\label{fig:ss2336}
\end{figure}

Fig.~\ref{fig:ss2336} manifests the case of initial state $\left|\boldsymbol{\psi}\left(t_{0}\right)\right\rangle=\left|23, 36\right\rangle$ with $T_{\mathrm{tot}}=6 T_{m}$.
Two bosons propagate toward each other until to occupy neighboring sites, then are forbidden to stay at the same site due to the strong interaction strength (named as interaction blockade), and finally yields a counter-propagating, see Fig.~\ref{fig:ss2336}(a).
After an evolved time $T_{\mathrm{tot}}=6 T_{m}$, the centroid remains almost constant with a zero centroid shifting in Fig.~\ref{fig:ss2336}(b).
Figs.~\ref{fig:ss2336}(c) and (d) respectively show the two-boson correlations at moments $t/T_m=0$ and $3$.
During the interaction blockade, two bosons behave similarly with Fig.~\ref{fig:ss2336}(d).
At last, two bosons nearly return to the initial position as Fig.~\ref{fig:ss2336}(c).

\section{Topological resonant tunnelings assisted by Bloch oscillations} \label{Chap5Sec4}

\begin{figure}[htp]
\center
\includegraphics[width=0.45\textwidth]{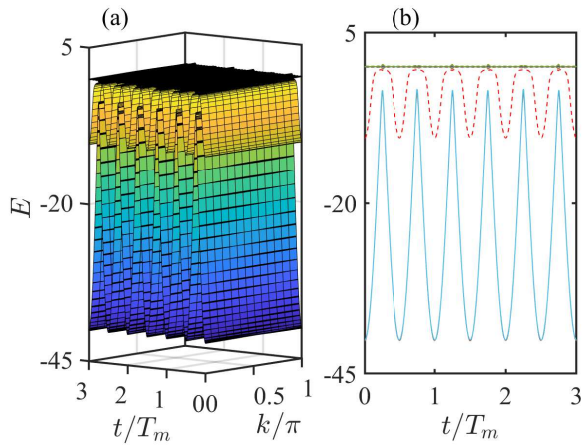}
\caption{The lowest three energy bands of two bosons.
(a) The 3D view of Bloch bands in the parameter space ($k$, $t$).
(b) The time evolution of energies $E(t)$ for Bloch states with $k=0$.
The parameters are chosen as $J=-1$, $\delta_{0}=0.8$, $\Delta_{0}=20$, $U=30$, $\omega=0.005$, $\phi_{0}=0$, $\omega_{F} / \omega=10 / 3$, $L_t=26$.}
\label{fig:resonantspectrum}
\end{figure}

Except for the topological transport of bound and scattering states, when $U \gg\left(J, \delta_{0}, \omega_{F}\right)$ and $\Delta_{0}>U / 2$ there will occur Topological resonant tunnelings as one of typical interaction effects.
The on-site potential difference between two neighboring sites will match the interaction strength four times over a period $T_m$, where the two bosons travel one by one rather than as a whole.
The effective single-particle model~\eqref{Chap5EffectiveModel} will be invalid owing to the important role of coupling between the subspaces $\mathcal{U}$ and $\mathcal{V}$.
Taking the Fock state as an example, once on-site potential difference between neighboring sites matches interaction, resonant tunneling occurs between state $|2\rangle_{j}$ and state $|1\rangle_{j}|1\rangle_{j+1}$ (or state $|1\rangle_{j-1}|1\rangle_{j}$)~\cite{2017PhysRevA95063630}.
When the parameters remain the same as Fig.~\ref{fig:boundspectrum} except for $\Delta_0=20$, the multiparticle Bloch bands are significantly changed, as shown in Fig.~\ref{fig:resonantspectrum}(a) with the lowest three energy bands.
Fig.~\ref{fig:resonantspectrum}(b) shows the change of energies for Bloch states with $k=0$, where a band marked with red dashed line is isolated from other bands to allow resonant tunnelings with band Chern number $C=3$.
The energy difference between the states $|1\rangle_{j}|1\rangle_{j+1}$ and $|2\rangle_{j}$ (or $|2\rangle_{j+1}$) almost vanishes at the energy avoided-crossings.
Within a modulation period $T_m$, four resonant tunnelings occur at four energy avoided-crossings.

\begin{figure}[htp]
\center
\includegraphics[width=0.45\textwidth]{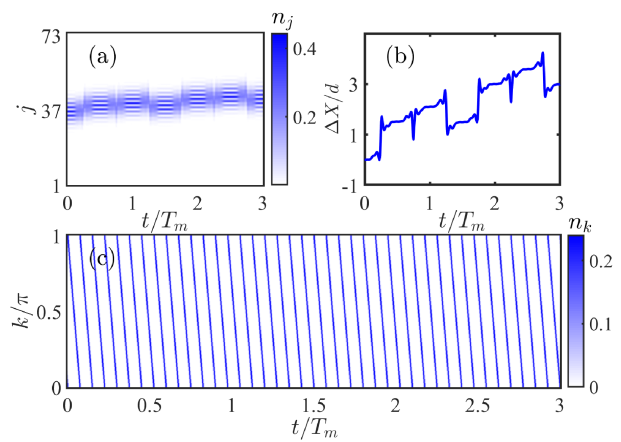}
\caption{Topological resonant tunnelings assisted by Bloch oscillations.
(a) and (b) are the time evolution of the particle density distribution in real space and the center-of-mass displacement, respectively.
(c) The time evolution of the particle density distribution in quasimomentum space.
The parameters are chosen as $J=-1$, $\delta_{0}=0.8$, $\Delta_{0}=20$, $U=30$, $\omega=0.005$, $\phi_{0}=0$, $\omega_{F} / \omega=10 / 3$ and $\sigma=5$.
The system size is $L_t=74$ and the evolution time is $T_{\mathrm{tot}}=q T_{m}=3 T_{m}$.}
\label{fig:resonanttransport}
\end{figure}

The multiparticle Bloch state with quasimomentum $k=0$ is constructed into Gaussian wave packet as the initial state.
The parameters are chosen as $J=-1$, $\delta_{0}=0.8$, $\Delta_{0}=20$, $U=30$, $\omega=0.005$, $\phi_{0}=0$, $\omega_{F} / \omega=10 / 3$ and $\sigma=5$.
The system size is $L_t=74$ and the evolution time is $T_{\mathrm{tot}}=q T_{m}=3 T_{m}$.
As shown in Fig.~\ref{fig:resonanttransport}(a), the initial Gaussian wave packet propagates locally in a certain range.
The wave-packet centroid moves up 2.99 cells in Fig.~\ref{fig:resonanttransport}(b), that is, $\Delta X (3T_{m})/ d=2.99\approx3$.
There is a slight deviation between the centroid displacement extracted from the wave-packet dynamics and the band Chern number, which is caused by the nonadiabatic effect in the time evolution due to the slight band gap at the energy avoided-crossings.
Based on the semiclassical formula~\eqref{Chap5Semiclassical}, we numerically calculate the reduced Chern number of multiparticle Bloch states with quasimomentum $k_0=0$, and obtain $C_{ \mathrm{red}}\left(3 T_{m}\right)=3$.
Both $\Delta X (3T_{m})/ d$ and $C_{\mathrm{red}}\left(3 T_{m}\right)$ are nearly the band Chern number.
Despite a very obvious difference in the real-space density distribution between topological resonant tunnelings and topological pumping of bound states, they both behave as linear scanning in quasimomentum space with a tilt $\omega_F$, see Fig.~\ref{fig:resonanttransport}(c) and Appendix~\ref{DDIQS}.
The reason is that, most of the time two bosons stay at the same lattice site, except near the avoided-crossing points.

\section{SUMMARY AND DISCUSSION} \label{Chap5Sec6}

We explore the correlated topological pumping by applying a tilted potential in an interacting Rice-Mele model.
Owing the role of tilt, the time variable not only provides an artificial dimension, but also carries out a uniform sampling in quasimomentum space.
Our scheme reduces the difficulty of initial-state preparation as well as the wave-packet dispersion, which any states prepared on the focused band can realize a dispersionless quantized transport.
The topological pumping of bound states assisted by Bloch oscillations is analytically explained via an effective single-particle model.
For scattering states, the fruitful topological transports can be engineered by choosing the two-particle initial positions.
The topological resonant tunnelings of two particles moving one by one are clarified when on-site potential difference between neighboring sites matches the interaction.

Our results are of great significance to topological pumping and can be generalized to spinor systems such as spin-1/2 bosons~\cite{2003PhysRevLett91010407,2016PhysRevLett117170405}, and other types of Thouless pumping such as nonlinear Thouless pumping~\cite{2021nature59663,2022PhysRevLett128154101}, Non-Abelian Thouless pumping~\cite{2021PhysRevA103063518,2022PhysRevLett128244302,2022NatPhySun}, fractional Thouless pumping~\cite{2022arxivfractional} and higher-order topological pumping~\cite{2022PhysRevB105195129}.
Take the spin component into account, it is possible to obtain a dispersionless topological spin transport with accessible initial states, which has potential application prospect for designing robust and flexible topological quantum devices, such as topological beam splitters.

\emph{Note added}: During the preparation of the manuscript, we became aware that Bloch oscillations contribute to the engineering of Floquet bands by applying a tilted linear potential~\cite{2022arxivFloquet}.

\begin{acknowledgments}
We acknowledge useful discussions with Bo Zhu and Zhoutao Lei. This work is supported by the NSFC (Grants No. 12025509, No. 11874434), the Key-Area Research and Development Program of GuangDong Province (Grants No. 2019B030330001), and the Science and Technology Program of Guangzhou (China) (Grants No. 201904020024).
Y.K. is partially supported by the NSFC (Grant No. 11904419).
W.L. is partially supported by the NSFC (Grant No. 12147108) and the
Fundamental Research Funds for the Central Universities, Sun Yat-Sen University (Grant No. 22qntd3101).
S.H. is partially supported by the NSFC (Grant No. 12104103).
L.Z. is partially supported by the Fundamental Research Funds for the Central Universities, Sun Yat-Sen University (Grant No. 22qntd3101).
\end{acknowledgments}

\appendix

\section{Bulk-boundary correspondence of multiparticle Bloch bands} \label{BBC}

\begin{figure*}[htp]
\center
\includegraphics[width=0.7\textwidth]{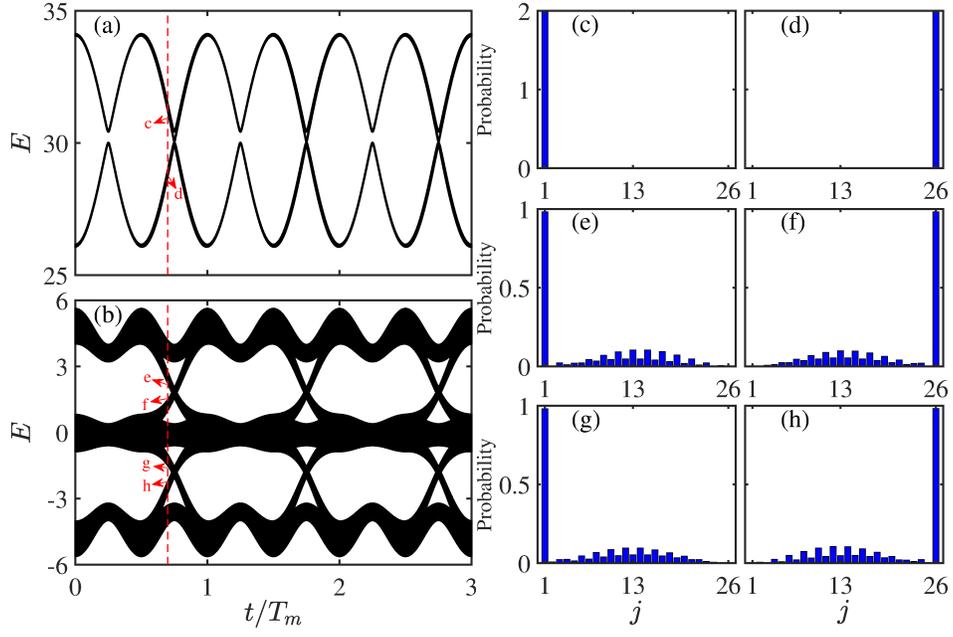}
\caption{Two-boson energy spectrum under the open boundary condition.
(a) and (b) are the energy spectra as a function of time $t$.
(c)-(h) are the density distributions of chosen states marked with c-h in red dashed lines in (a) and (b).
The parameters are chosen as $J=-1$, $\delta_{0}=0.8$, $\Delta_{0}=2$, $U=30$, $\omega=0.005$, $\phi_{0}=0$, $\omega_{F} / \omega=10 / 3$, $L_t=26$.}
\label{fig:bbc}
\end{figure*}

The bulk-boundary correspondence is a key feature in topological band theory, that is, nontrivial band topology is associated with the existence of topological edge states
at open boundaries~\cite{1993PhysRevB4811851,1993PhysRevLett713697}.
Compared with multiparticle Bloch band in Fig.~\ref{fig:boundspectrum}(a) in the main text, under the open boundary condition isolated states appear at the energy gap in Figs.~\ref{fig:bbc}(a) and (b).
For the chosen parameter marked with red dashed lines in Figs.~\ref{fig:bbc}(a) and (b), we respectively extract states c-h to compute the density distributions: left (right) bound-boson edge state in Fig.~\ref{fig:bbc}(c) [Fig.~\ref{fig:bbc}(d)], left (right) single-boson edge states in Figs.~\ref{fig:bbc}(e) and (g) [Figs.~\ref{fig:bbc}(f) and (h)].

\section{Density distribution in quasimomentum space} \label{DDIQS}
\begin{figure}[htp]
\center
\includegraphics[width=0.4\textwidth]{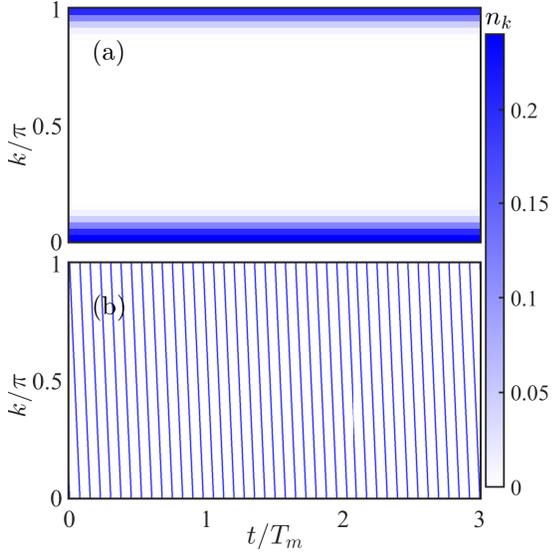}
\caption{The time evolution of particle density distribution in quasimomentum space for different values of $\omega_F$: (a) $\omega_F=0$ and (b) $\omega_F=10 \omega/ 3$.
The other parameters are chosen as $J=-1$, $\delta_{0}=0.8$, $\Delta_{0}=2$, $U=30$, $\omega=0.005$, $\phi_{0}=0$, and $\sigma=5$.
The system size is $L_t=74$ and evolution time is $T_{\mathrm{tot}}=q T_{m}=3 T_{m}$.}
\label{fig:dms}
\end{figure}
In the presence of a tilt, it is challenging to extract the period of the Bloch oscillations in real space.
By transferring the evolved states in Figs.~\ref{fig:boundtransport}(q) and (w) in the main text to the $k$-space representation, the average quasimomentum of the wave packet respectively remains unchanged and linearly scans the Brillouin zone as time evolves, as shown in Figs.~\ref{fig:dms}(a) and (b).
Fig.~\ref{fig:dms}(b) manifests, when the quasimomentum hits the boundary $k=0$ in the Brillouin zone, it immediately transitions to another boundary $k=\pi $.
The number of Bloch oscillations can be clearly extracted by counting the number for scanning the whole Brillouin zone of the average quasimomentum.
We find that the wave packet undergoes fractional Bloch oscillations with a period half of the single-particle Bloch-oscillation period $T_{B}=2\pi / (d\omega_{F})$, which means that the effective tilt experienced by two strongly interacting bosons turns to be $2\omega_F$.
Naturally, the time evolution of particle density distribution in quasimomentum space can be used to determine the value of the tilt, which are potential applications in the measurement of gradient magnetic field or gravitational field.

\section{The relationship between reduced Chern number and Chern number} \label{appendixRC}

We prove that the reduced Chern number of bound-state bands $C_{m, \mathrm{red}}$ is independent of the value of quasimomentum, and build its relationship with Chern number.
Based on the analytical expression of Berry curvature~\eqref{Chap5BerryCurvature1} in the main text, it can be seen that in the absence of tilted potential the Berry curvature turns to be
\begin{widetext}
\begin{equation}
\mathcal{F}^{0}_{\pm}(k, t)=32 \frac{J \delta_{0} \omega \Delta_{0}}{U} \frac{\left\{J^{2}+\delta_{0}^{2} \sin ^{2}[\phi(t)]\right\}}{U} \frac{1-\cos ^{2}[\phi(t)] \cos ^{2}\left(k\right)}{\left[\varepsilon^{0}_{\pm}(k, t)\right]^{3}},
\end{equation}
where
\begin{equation}\label{Chap5Withouttilt}
\varepsilon_{\pm}^{0}(k, t)
=\pm \sqrt{\left(4 \frac{\left\{J^{2}+\delta_{0}^{2} \sin ^{2}[\phi(t)]\right\}}{U} \cos k\right)^{2}+\left(4 \frac{\left\{2 J \delta_{0} \sin [\phi(t)]\right\}}{U} \sin k\right)^{2}+\left(2 \Delta_{0} \cos [\phi(t)]\right)^{2}}+\mathcal{C}
\end{equation}
is the eigenvalue of bound-state bands for $\omega_F=0$.
There exist $\varepsilon_{\pm}(k, t)=\varepsilon_{\pm}^{0}\left(k-2\omega_{F} t, t\right)$ and $\mathcal{F}_{\pm}(k, t)=\mathcal{F}_{\pm}^{0}\left(k-2\omega_{F} t, t\right)$.
As $k$ goes to $k+\Delta k$ and $t$ goes to $t+\Delta k/(2\omega_F)$, the Berry curvature yields
\begin{equation}\label{Chap5BerryCurvature2}
\begin{array}{l}
\mathcal{F}_{\pm}\left(k+\Delta k, t+\Delta k / 2 \omega_{F}\right) \\
=\frac{\pm 32 J \delta_{0} \omega \Delta_{0}\left\{J^{2}+\delta_{0}^{2} \sin ^{2}\left[\phi(t)+\phi_s\right]\right\}\left\{1-\cos ^{2}\left[\phi(t)+\phi_s\right] \cos ^{2}\left(k-2 \omega_{F} t\right)\right\}}{U^{2}\left\{\left[4 \frac{\left\{J^{2}+\delta_{0}^{2} \sin ^{2}\left[\phi(t)+\phi_s\right]\right\}}{U} \cos \left(k-2 \omega_{F} t\right)\right]^2+\left[4 \frac{\left\{2 J \delta_{0} \sin \left[\phi(t)+\phi_s\right]\right\}}{U} \sin \left(k-2 \omega_{F} t\right)\right]^2+\left[2 \Delta_{0} \cos \left[\phi(t)+\phi_s\right]\right]^{2}\right\}^{\frac{3}{2}}}\\
\approx\frac{\pm 32 J \delta_{0} \omega \Delta_{0}\left\{J^{2}+\delta_{0}^{2} \sin ^{2}[\phi(t)]\right\}\left\{1-\cos ^{2}[\phi(t)] \cos ^{2}\left(k-2 \omega_{F} t\right)\right\}}{U^{2}\left\{\left[4 \frac{\left\{J^{2}+\delta_{0}^{2} \sin ^{2}[\phi(t)]\right\}}{U} \cos \left(k-2 \omega_{F} t\right)\right]^2+\left[4 \frac{\left\{2 J \delta_{0} \sin [\phi(t)]\right\}}{U} \sin \left(k-2 \omega_{F} t\right)\right]^2+\left[2 \Delta_{0} \cos \phi(t)\right]^{2}\right\}^{\frac{3}{2}}}\\
=\mathcal{F}_{\pm}(k, t)
\end{array}
\end{equation}
\end{widetext}
with $\phi_s=\frac{\omega}{2 \omega_{F}} \Delta k$.
When $\omega_{F} / \omega \rightarrow \infty$, we have $\phi_s=0$ and the relation~\eqref{Chap5BerryCurvature2} becomes exact enough.
In the following derivation, we assume that the system respects $\omega_{F} / \omega \rightarrow \infty$.
Since Berry curvature~\eqref{Chap5BerryCurvature2} is a periodic function with period $qT_m$ in time domain, the one-dimensional time integral of Berry curvature ~\eqref{Chap5BerryCurvature2} follows
\begin{equation}\label{Chap5BerryCurvature3}
\begin{array}{l}
\int_{0}^{q T_{m}} \mathcal{F}_{\pm}\left(k+\Delta k, t+\Delta k / 2\omega_{F}\right) d t \\
=\int_{\Delta k / 2\omega_{F}}^{q T_{m}+\Delta k / 2\omega_{F}} \mathcal{F}_{\pm}(k+\Delta k, t) d t \\
=\int_{0}^{q T_{m}} \mathcal{F}_{\pm}(k+\Delta k, t) d t.
\end{array}
\end{equation}
According to Eqs.~\eqref{Chap5BerryCurvature2} and~\eqref{Chap5BerryCurvature3}, we have
\begin{equation}
\begin{aligned}
C_{m, \mathrm{red}} &=\frac{1}{d} \int_{0}^{q T_{m}} \mathcal{F}_{\pm}(k, t) d t \\
&=\frac{1}{d} \int_{0}^{q T_{m}} \mathcal{F}_{\pm}(k+\Delta k, t) d t .
\end{aligned}
\end{equation}
It means that the reduced Chern number of bound-state bands $C_{m,\text{red}}$ is independent of the value of quasimomentum $k$ with
\begin{equation}
C_{m, \mathrm{red}}(k+\Delta k)=C_{m, \mathrm{red}}(k).
\end{equation}
Thus we can average the reduced Chern number over the Brillouin zone without affecting the final results as
\begin{equation}
\begin{aligned}
C_{m, \mathrm{red}} \left(q T_{m}\right)& = \frac{1}{2 \pi} \int_{0}^{q T_{m}} \int_{-\pi / d}^{\pi / d} \mathcal{F}_{m}^{0}\left(k-2\omega_{F} t, t\right) d t d k \\
&=\frac{1}{2 \pi} \int_{0}^{qT_{m}} \int_{-\pi / d}^{\pi / d} \mathcal{F}_{m}^{0}(k, t) d t d k=C_{m}.
\end{aligned}
\end{equation}
Different from the Chern number obtained by a two-dimensional integral, the reduced Chern number is a one-dimensional integral over time.
The reason is that in the presence of a tilt, all quasimomentum values are uniformly sampled over a time cycle.
The quasimomentum sampling can be understood as a valid ergodic behavior in the Brillouin zone, such that the one-dimensional time integral is independent of the specific value of the initial quasimomentum $k_0$.

%


\end{document}